\documentstyle[twoside,fleqn,espcrc2]{article}

\def\KL{$K_{L}$}
\def\qm{$\pm$}
\def\pizero{\pi^{0}}
\def\piz{$\pi^{0}$}
\def\rarrow{\rightarrow}

\def\etal{{\sl et al.}}
\def\tan2thc{{\rm tan}^2\theta_C}
\def\costhk{{\rm cos}\theta_K}
\def\sinthk{{\rm sin}\theta_K}
\def\BR#1{{\cal B}(#1\nu_\tau)}
\def\deg{$^\circ$}

\newcommand{\AmS}{{\protect\the\textfont2
  A\kern-.1667em\lower.5ex\hbox{M}\kern-.125emS}}

\title{Experimental Summary on Hadronic Decays: A TAU98 Review
\thanks{Invited talk at {\it The Fifth Workshop on Tau Lepton 
        Physics,} Santander, Spain, 14-18 September 1998.}}

\author{Brian K. Heltsley\address{Wilson Synchrotron Laboratory,
        Cornell University \\ 
        Ithaca, NY 14853   USA\\
        bkh@mail.lns.cornell.edu } }
       
\begin{document}

\begin{abstract}
Selected results on hadronic decays of the tau lepton from the 
TAU98 Workshop are reviewed. A comprehensive picture emerges
for strange particle branching fractions, and exploration of 
resonant substructure of
both strange and non-strange decays is seen to have matured substantially.
\end{abstract}

\maketitle

\section{Introduction}

\thispagestyle{myheadings}
\markright{{\rm CLNS~98/1590\ \ November 1998}}

    Unique among the leptons, the tau is massive enough to decay 
semi-hadronically. Many hadronic branching fractions are 
predicted with precision in the Standard Model, motivating
correspondingly accurate measurements as probes of new physics.
Tau decays also furnish a remarkably potent production 
source for mesons, the properties of which are then available
for study.

    Although the internal inconsistencies and disagreements between 
measured and predicted 
branching fractions have diminished in the past several years,
the process of rate measurements for every conceivable mode 
and unrelenting improvement of precision continues. The larger branching
fractions are measured with relative errors below one percent, signals for 
rare decays are seen at the $10^{-4}$ level, and forbidden decay rate 
upper limits approach $10^{-6}$. Anticipating the completion of the LEP~I
and CLEO~II programs, experimenters are exploiting the
large datasets and pursuing improved understanding of the detectors 
and better analysis techniques to further study hadronic tau decays.

   In earlier Workshops the experimental emphasis initially
focussed on putting \piz\ detection on par with that of charged
particles for modes with one or multiple \piz's; 
strange particle identification efforts moved from detached vertex
finding for $K_S$, to $dE/dx$ or RICH calibration for $K^-$, to
calorimetry for $K_L$. These techniques have been refined
while datasets have grown, allowing more challenging modes
involving {\it multiple} kaons and/or {\it multiple} \piz's to be measured.

  In order to make quantitative conclusions from the large number of new
and old measurements, the common practice of forming weighted averages of all 
branching fractions is followed here. For each mode, the ``New World Average'' 
(NWA) weights each of $n$ measurements by its inverse-square total error, 
which is the quadrature sum of statistical and systematic contributions. 
The NWA error is computed as the inverse of the quadrature sum of the 
inverse total errors, scaled up by the Particle Data Group \cite{PDG98} 
scale factor $max(1,S)$, where $S=\sqrt{\chi^2/(n-1)}$. For 
this review, I have omitted any entries that, by virtue of their errors 
relative to others, contribute less than 5\% of the weight in the NWA. 
PDG98 \cite{PDG98} contains the complete 
listings. These NWA's only are only approximate for a number 
of reasons: many newer measurements are preliminary;
some systematic errors may be correlated and therefore should not be 
combined in quadrature; biased weighting can occur for statistics-limited
measurements. In addition, measurements of the same quantity
sometimes make different assumptions for cross-feed branching ratios
or resonant content, and hence should be converted to a common basis
before being averaged. Finally, a set of measurements can be internally 
inconsistent (have a scale factor $S$ significantly larger than unity), 
in which case the NWA does not
have a clear interpretation.

  The scope of this summary does not allow for a complete or comprehensive 
compilation of every new result since TAU96; in particular, measurements
inclusive of neutrals have been largely neglected. The reader is referred to 
the other presentations at this conference for up-to-date information on
V/A spectral functions and the determination of $\alpha_s$ in tau 
decay \cite{Menke,Hoecker}, 3$\pi$ 
sub-structure \cite{Schmidtler}, and the impact of tau decays on hadronic 
contributions to $(g-2)_\mu$ and $\alpha_{QED}$ \cite{Davier}.

\section{PDG98 branching fractions \& fit}

\thispagestyle{myheadings}
\markright{ }

  There have been several hadronic branching fraction results published
since TAU96, so consulting PDG98 \cite{PDG98} is quite worthwhile.
A systematic compilation with averages by specific decay appears 
along with a global fit. In the fit, measured decays are expressed in
terms of a set of exclusive basis decay modes, and the sum of 
the basis mode branching fraction is constrained to unity. The key 
observation by the PDG for tau decays in this edition is that there was
only one basis mode ($\BR{\pi}$) that changed from PDG96 \cite{PDG96} by
more than one standard deviation. Here the most important feature of tau 
branching fractions is one of gradually improving precision, internal 
consistency, and stability of results. ``Deficits'' or ``problems'' in 
topological branching fraction vs. itemized decay mode sums, or
significant internal disagreements in individual modes are mostly 
``old news''. The usually more precise recent measurements tend to be 
consistent with each other, and the troubles of the past seem to have been
caused by systematic errors in the older, (now) less precise measurements.

  Nevertheless it may be useful to probe the limitations of the
PDG98 conclusions. The global fitting procedure does not handle
explicit correlations between measured branching fractions in 
different modes, even though such correlations have sometimes been given along
with the measurement results. Six measured decay modes totaling 
(0.15\qm0.05)\% are explicitly ignored in the fit because they 
cannot be expressed 
in terms of the basis modes. A special unofficial fit with a catch-all
unconstrained ``dummy'' decay mode results in this mode being
attributed with (0.50$^{+0.40}_{-0.29}$)\% branching 
fraction \cite{Moenig}. Hence the basis mode averages sum to less than unity, 
and in the official global
fit, some branching fractions are pulled noticeably higher to 
compensate. There may be unmeasured modes with branching fraction sum 
at the $\sim$0.5\% level, or there may be problems with some of the 
measurements, or this may be a statistical fluctuation. However, there 
are other odd features to the global fit that have led the PDG to plan 
for a new fitter \cite{Moenig}, one that will better address the stated 
goal of PDG to attain 0.1\% internal consistency.

\section{CVC Tests}

\begin{figure}[htb]
\vskip1.9in
\caption{Branching fractions for $\tau^-\rarrow h^-\pizero\nu_\tau$.}
\label{fig:fhpi0nu}
\includegraphics{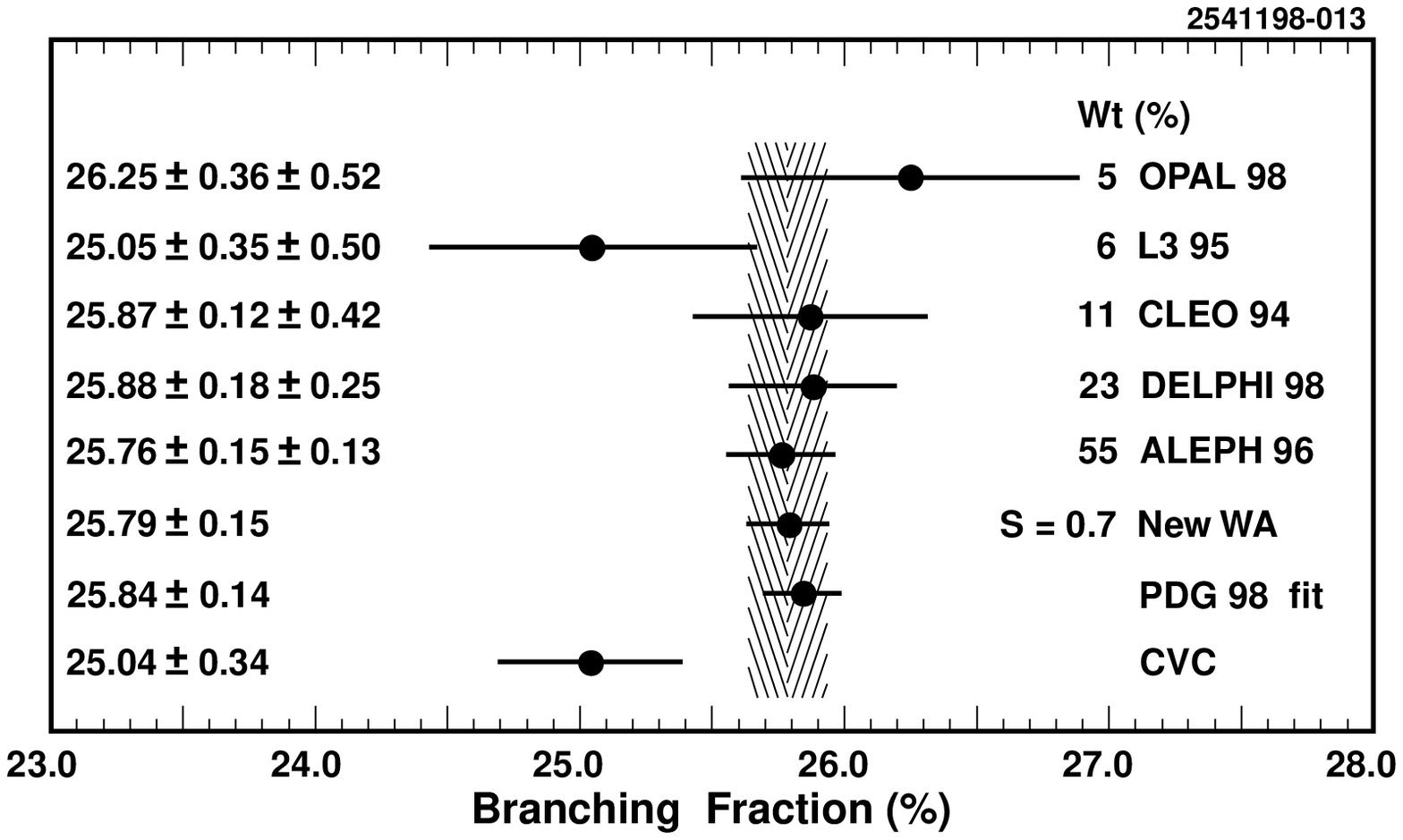}
\end{figure}

  Those tau decays that proceed through the vector current are subject to 
tests of the Conserved Vector Current and therefore comparison with 
electron-positron annihilation to the same final states. With the increase 
in amount and quality of tau data, in recent years tests of CVC have become 
limited by both the lack of comparably precise data from low energy $e^+e^-$ 
colliders and the subtlety of systematic effects such as radiative
corrections, separation of isoscalar components, and interference. 
This situation is improving somewhat with recent data from the CMD2
and SND experiments operating at the VEPP-2M collider at BINP in
Novosibirsk. A compilation \cite{Eidelman} of the measured versus predicted
branching fractions for vector-current modes shows a 2.4 standard
deviation excess in the overall sum of tau rates compared with CVC
prediction, with the excess concentrated in the two modes
with largest branching fraction: $h\pizero$ and $3h\pizero$ 
(see Figs.~\ref{fig:fhpi0nu}, \ref{fig:f3hpi0nu}). 
At this point it is too soon to tell whether this is
significant given the experimental and theoretical subtleties involved.
Further work is necessary, and improved data are expected on both 
tau and $e^+e^-$ sides.

\begin{figure}[htb]
\vskip2.05in
\caption{Branching fractions for $\tau^-\rarrow h^-\nu_\tau$.}
\label{fig:fhnu}
\includegraphics{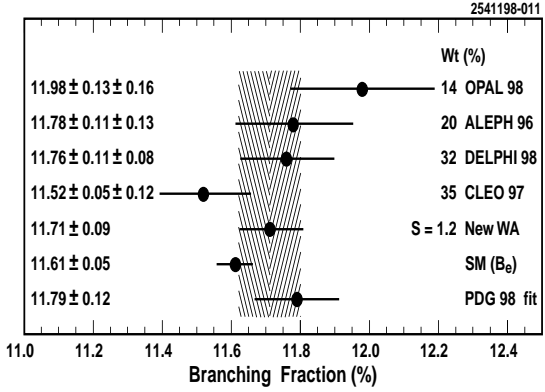}
\end{figure}

\section{Non-strange Decays}

\begin{figure}[htb]
\vskip1.8in
\caption{Branching fractions for $\tau^-\rarrow h^-2\pizero\nu_\tau$.}
\label{fig:fh2pi0nu}
\includegraphics{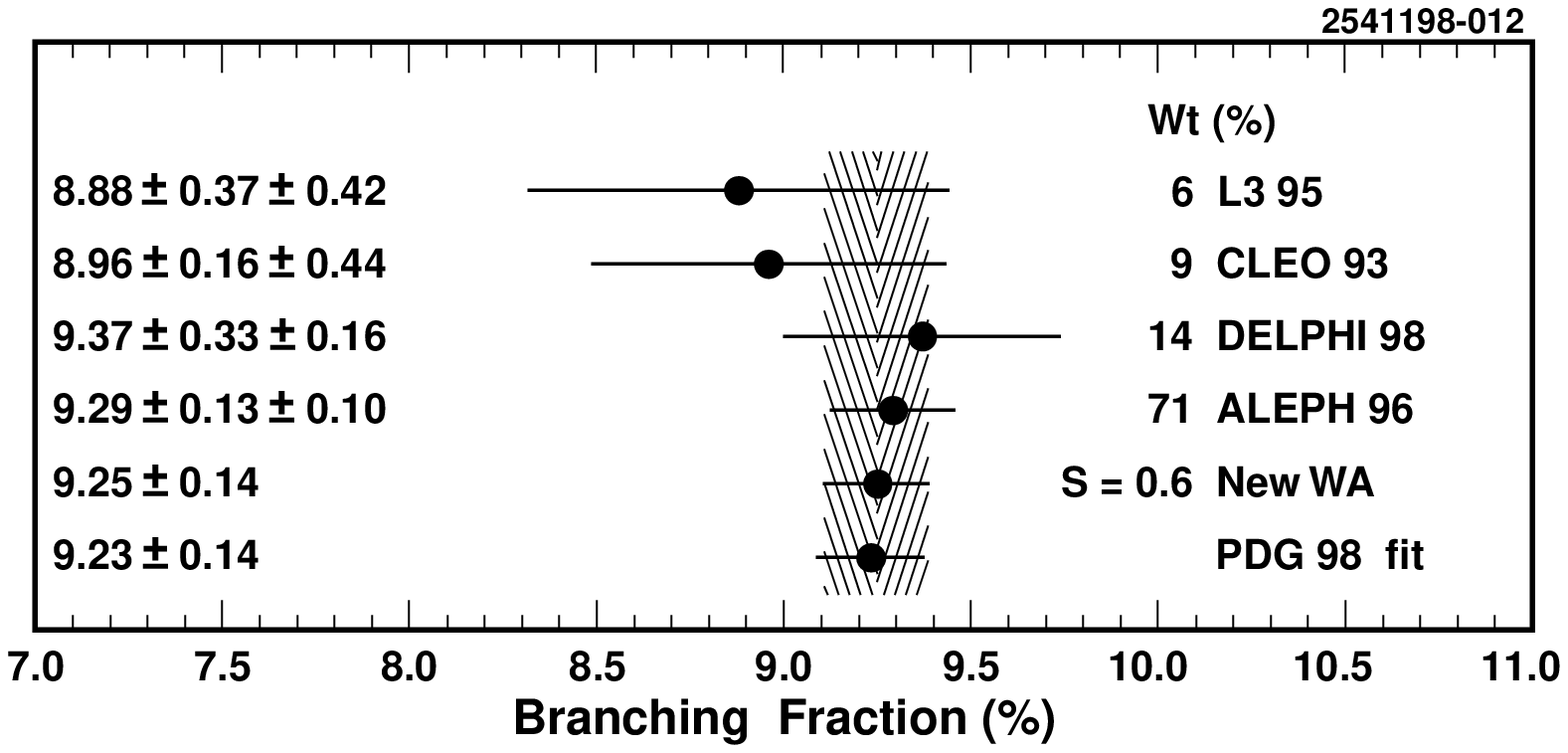}
\end{figure}

\begin{figure}[htb]
\vskip1.45in
\caption{Branching fractions for $\tau^-\rarrow 3h^\pm\nu_\tau$.}
\label{fig:f3hnu}
\includegraphics{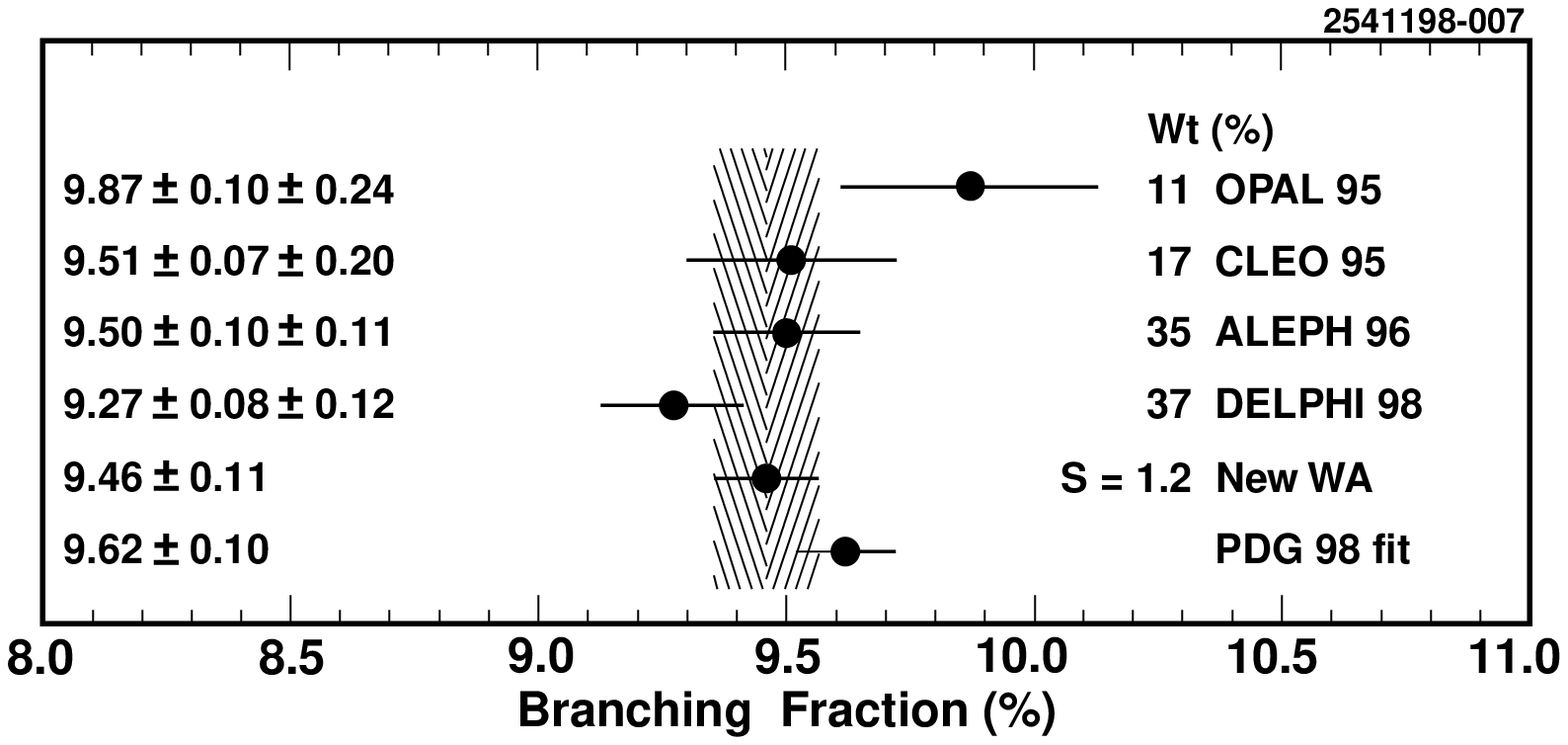}
\end{figure}

\begin{figure}[htb]
\vskip2.2in
\caption{Branching fractions for $\tau^-\rarrow 5h^\pm\nu_\tau$.}
\label{fig:f5hnu}
\includegraphics{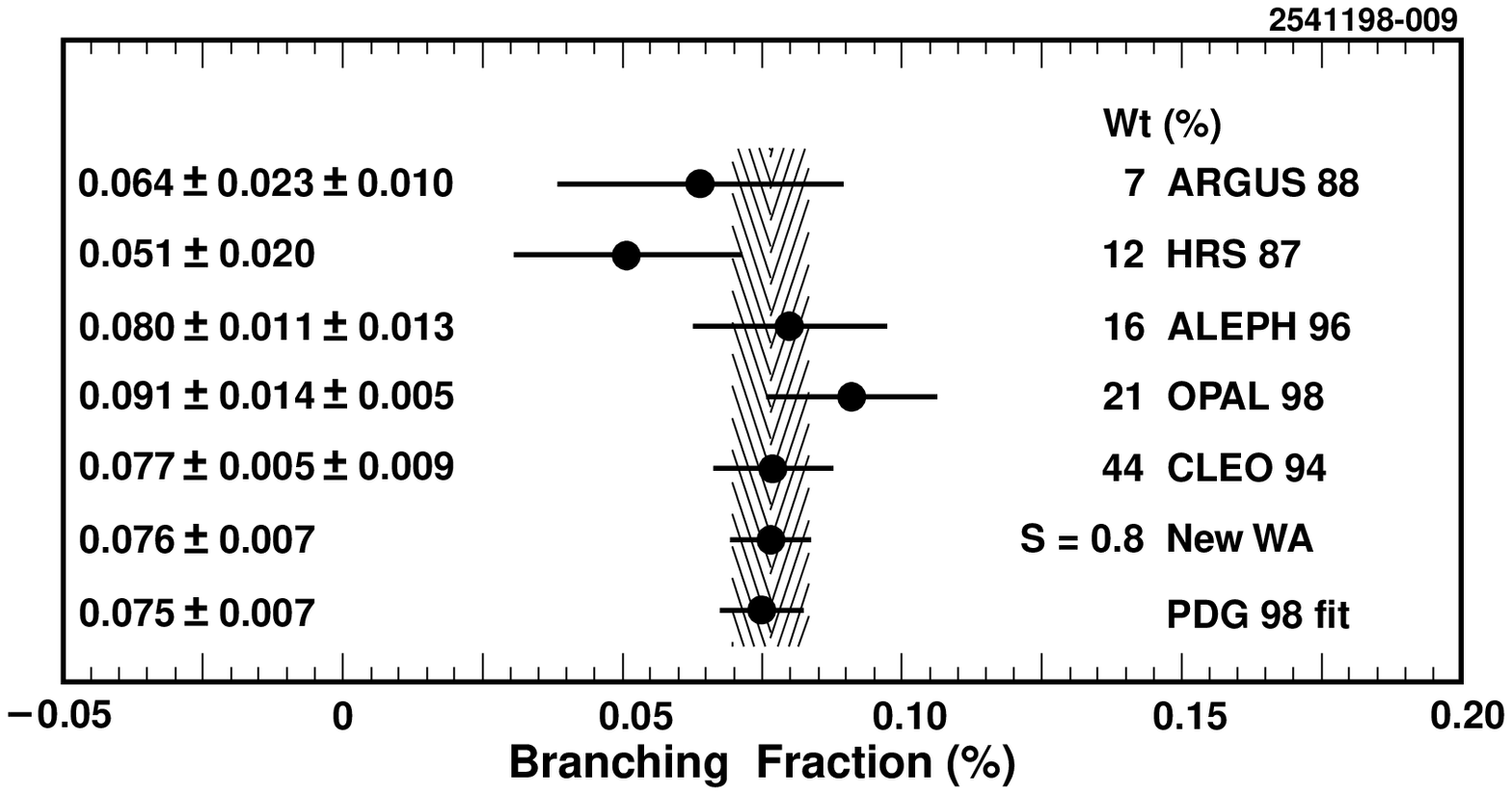}
\end{figure}

\begin{figure}[htb]
\vskip1.6in
\caption{Branching fractions for $\tau^-\rarrow 5h^\pm\pizero\nu_\tau$.} 
\label{fig:f5hpi0nu}
\includegraphics{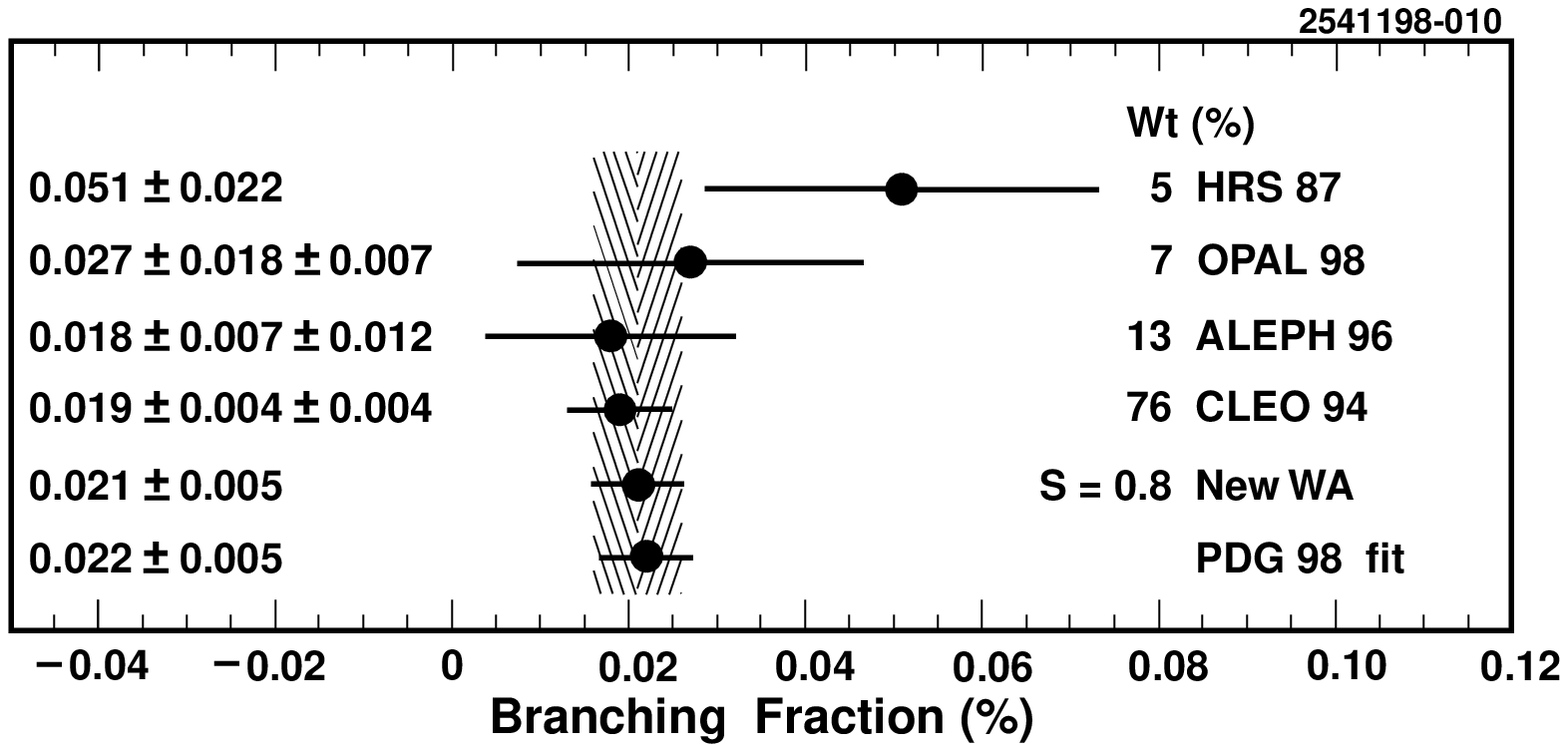}
\end{figure}

\begin{figure}[htb]
\vskip1.8in
\caption{Branching fractions for $\tau^-\rarrow 3h\pizero\nu_\tau$.}
\label{fig:f3hpi0nu}
\includegraphics{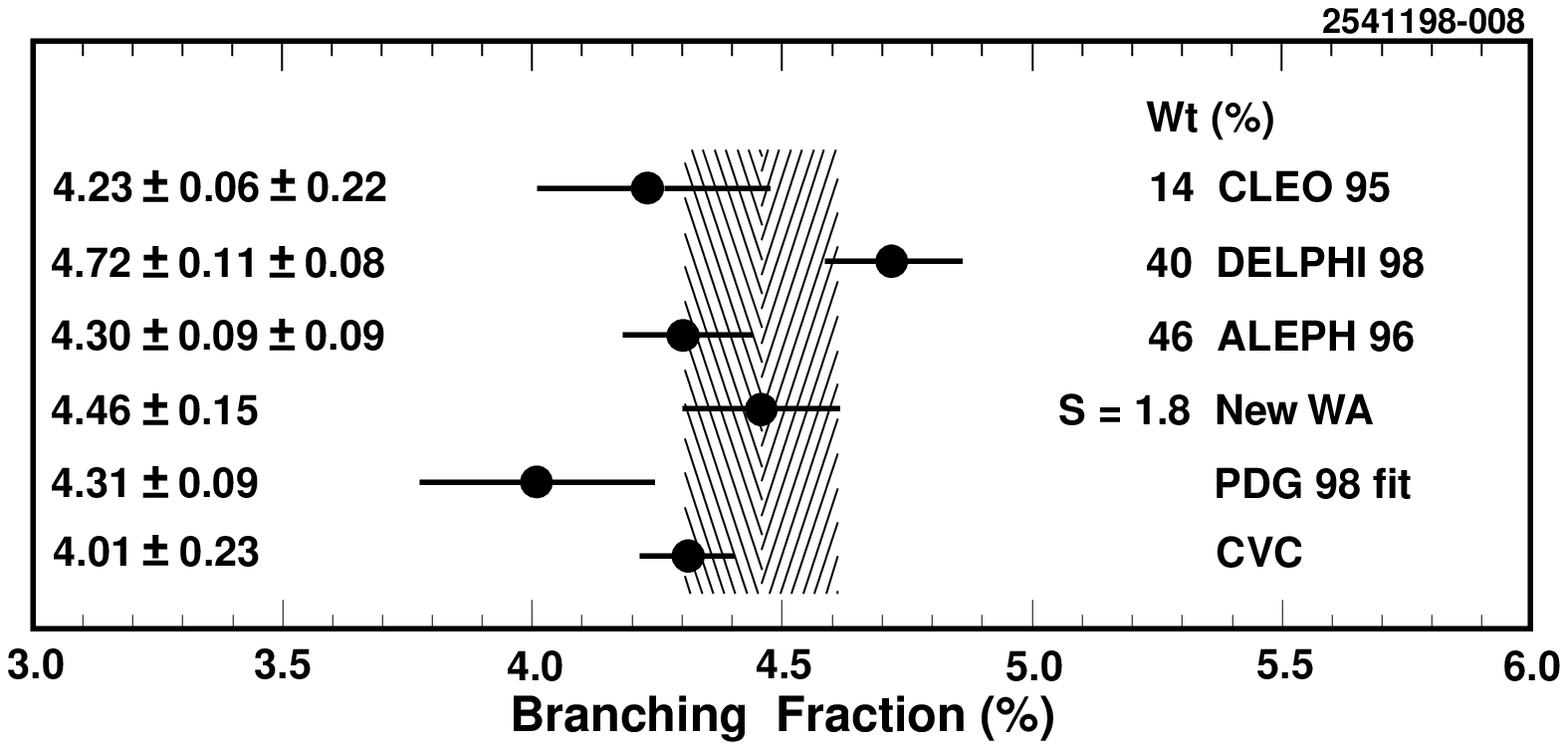}
\end{figure}

  OPAL \cite{OPAL1prong,OPAL5prong} and DELPHI \cite{Lopez-Garcia} have
new branching fraction results in several 
non-strange modes made possible by new understanding of how \piz's can 
be detected in their calorimeters. For $\BR{h}$, $\BR{h\pizero}$, 
$\BR{h2\pizero}$, $\BR{3h^\pm}$, $\BR{5h^\pm}$, and $\BR{5h^\pm\pizero}$, 
as shown in Fig.~\ref{fig:fhnu}-\ref{fig:f5hpi0nu}, 
the new entries are compatible with and comparable in precision to 
previous measurements. However, the preliminary DELPHI measurement of 
$\BR{3h^\pm\pizero}$ (Fig.~\ref{fig:f3hpi0nu}) is 2.5 standard
deviations higher than the PDG98 fit value. It is also 2.6 standard
deviations from the CVC prediction for 
$\BR{3\pi^\pm\pizero}=(3.90\pm0.22)\%$ \cite{Eidelman} augmented by 
the NWA's for $\BR{K^-\pi^+\pi^-\pizero}+\BR{K^-K^+\pi^-\pizero}$ 
(Figs.~\ref{fig:fk2pipi0nu} and \ref{fig:fkkpipi0nu}).

  There have also been new results on other high multiplicity final states.
CLEO finalized their TAU96 analysis \cite{Shelkov96} on 
$\tau\to f_1\pi^-\nu_\tau$ with a publication \cite{CLEOf1pi}, reporting 
$\BR{3\pi^\pm\eta}=(3.4^{+0.6}_{-0.5}\pm0.6)\times 10^{-4}$,
$\BR{\pi^-2\pizero\eta}=(1.4\pm0.6\pm0.3)\times 10^{-4}$, and
$\BR{f_1\pi^-}=(5.8^{+1.4}_{-1.3}\pm1.8)\times 10^{-4}$.
A new measurement \cite{Barberis} of the $f_1(1285)\to\eta\pi\pi$ 
branching fraction as (52.8\qm3.8\qm2.5)\% allows reduction of
the CLEO systematic error, resulting in 
$\BR{f_1\pi^-}=(5.9^{+1.4}_{-1.3}\pm1.2)\times 10^{-4}$.
 In a related mode, CLEO reports
$\BR{3\pi^\pm3\pizero}=(2.85\pm0.56\pm0.51)\times10^{-4}$,
finding that it is saturated by the above $3\pi\eta$ modes and
the (new) $\pi^-2\pizero\omega$ final state with rate 
$\BR{\pi^-2\pizero\omega}=(1.89^{+0.74}_{-0.67}\pm0.40)\times10^{-4}$
 \cite{Gan,CLEO3pi3pi0}. The older ALEPH measurement of
$\BR{3\pi^\pm\ge3\pizero}=(11\pm4\pm5)\times10^{-4}$ \cite{PDG98}
is larger but consistent with the new CLEO result; although it
should be noted that $\BR{3\pi^\pm\ge4\pizero}$ has not yet been
measured directly. 
Measured six pion branching fractions
(for $3\pi^\pm3\pizero$ and $5\pi^\pm\pizero$) are found to be
consistent with isospin symmetry once the axial-vector
$(3\pi)^-\eta$ contributions have been subtracted \cite{Gan,CLEO3pi3pi0}. 
Finally, new upper limits on tau decays with
seven charged particles plus neutrals have been made by OPAL \cite{OPAL7pi}
(1.8$\times 10^{-5}$ at 95\%CL) and CLEO \cite{CLEO7pi}
(2.4$\times 10^{-6}$ at 90\%CL).

\section{Modes involving strange particles}

  Much of the focus of new work on hadronic decays focused on
modes with strange particles. ALEPH presented a comprehensive
analysis \cite{Chen} of 27 different modes, almost always measuring
both $K_S$ and $K_L$ for modes with $K^0$'s.

\begin{figure}[htb]
\vskip1.35in
\caption{Branching fractions for $\tau^-\rarrow K^-\nu_\tau$.}
\label{fig:fknu}
\includegraphics{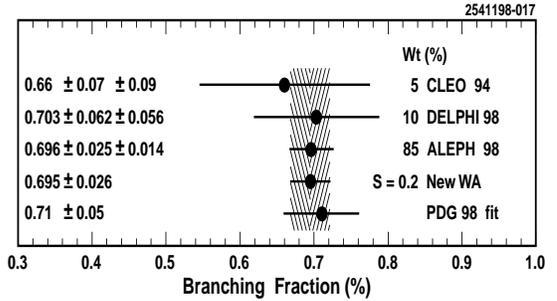}
\end{figure}

\subsection{$\tau\rarrow K^-\nu_\tau$}

  Both ALEPH \cite{Chen,ALEPHKaon1pr} and DELPHI \cite{Andreazza,DELPHIKaon} 
have new branching ratios for $\BR{K}$; they are consistent with previous
measurements, and ALEPH halves their previous (smallest) error, which 
improves the world average accordingly, as shown in Fig.~\ref{fig:fknu}. 
The ALEPH analysis uses this improved precision to test $\tau -\mu$ 
universality ($g_\tau/g_\mu=0.987\pm 0.021$) \cite{ALEPHKaonSpc} by
comparing the decay rate to that of muonic kaon decay, and to determine 
the kaon decay constant, assuming lepton universality, as 
$f_K = (111.5\pm 0.9\pm 2.3$)~MeV.

\begin{figure}[htb]
\vskip 2.0in
\caption{Branching fractions for combined $\tau^-\rarrow (K\pi)^-\nu_\tau$,
assuming an isospin $1\over 2$ parent such as $K^*(892)$ and/or $K^*(1410)$. 
 }
\label{fig:fkstnu}
\includegraphics{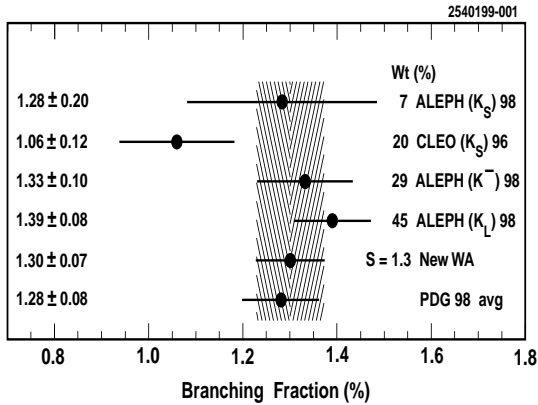}
\end{figure}

\subsection{$\tau\rarrow (K\pi)^-\nu_\tau$}

  ALEPH \cite{Chen,ALEPHKaon1pr,ALEPHKaonSpc} 
presented new results for 
$K^-\pizero$, $K_S\pi^-$, and $K_L\pi^-$ final states. 
Fig.~\ref{fig:fkstnu} shows the $(K\pi)^-$ branching fraction, labeled 
by specific final state, assuming an isospin $1\over 2$ parent. 
The CLEO result remains about two standard deviations lower than 
the more precise combined ALEPH rate.

 ALEPH also fits the $K\pi$ spectrum for the presence of $K^*(1410)$ 
in addition to $K^*(892)$, and for interference between the two final
states. They obtain evidence for $K^*(1410)$ at the one to three
standard deviation level, depending upon which parameters are fixed or 
allowed to float in the fit and which decay mode spectra are included.
The $K^*(1410)$ admixture strength $\beta$ is determined to be
$\sim$$-$0.1, similar to that measured for $\rho^\prime(1450)$
in $\tau\to\pi^-\pizero\nu_\tau$ \cite{ALEPHrho,Urheim96}. The net branching 
fraction is measured as $\BR{K^*(1410)}$=(0.15$^{+0.14}_{-0.10}$)\%.
As $K^*(1410)$ decays 93\% of the time to $(K\pi\pi)^-$,
this measurement yields the vector component background
to the dominantly axial-vector $(K\pi\pi)^-$ final states.

\subsection{$\tau\rarrow (K\pi\pi)^-\nu_\tau$}

   In analogy with tau decays to non-strange $(3\pi)^-$ states, the 
production of $(K\pi\pi)^-$ is expected to occur through the axial-vector 
mesons $K_1(1270)$ and 
$K_1(1400)$ \cite{Ronan,Suzuki93,Suzuki94,Lipkin,Heltsley94}. 
However, immediate non-resonant fragmentation to $(K^*\pi)^-$ or
$(K\rho)^-$ at the 
pointlike weak vertex, or production through the strange vector meson 
$K^*(1410)$ (see previous section), are also possible.

    Let $K_A$ and $K_B$ denote the triplet $^3P_1$ 
and singlet $^1P_1$ spin eigenstates, respectively, of $(\bar{u}s)$, the 
strange analogs of the $\bar{u}d$ states, $a_1(1260)$ and $b_1(1235)$. 
Production amplitudes of $b_1$ and $K_B$ in tau decay are suppressed 
in the SM as 
second-class currents by
$$\sim{{\vert m_u - m_{d,s}\vert}\over{m_u + m_{d,s}}}.$$
\noindent This factor is quite small for $m_d$, accounting for the 
non-observation 
of $b_1$ in tau decays. In contrast, the large strange quark mass induces 
substantial 
$SU(3)_f$ symmetry-breaking, allowing non-zero $K_B$ production in tau decay.

   The strange axial-vector mass eigenstates are $K_1(1270)$ and 
$K_1(1400)$. $K_1(1270)$ decays to $K\rho$, $K^*_0(1430)\pi$, $K^*\pi$, and 
$K\omega$; $K_1(1400)$ decays almost exclusively to $K^*\pi$. Since $K_A$ 
and $K_B$ are both predicted by $SU(3)$ to decay equally to $K^*\pi$ and 
$K\rho$, and since the $K_1$ states do not, the $K_1$ resonances must 
correspond to mixtures of $K_A$ and $K_B$ in comparable amounts with 
opposite-sign phases. One can parametrize this mixing in terms of an angle 
$\theta_K$:
$$ K_1(1400) = K_A~\costhk - K_B~\sinthk $$
$$ K_1(1270) = K_A~\sinthk + K_B~\costhk $$
\noindent The previously measured properties of the $K_1$ mesons place some
restrictions on the allowed values of $\theta_K$ \cite{Suzuki93,Suzuki94}. The
relative amounts of $K_1(1270)$ and $K_1(1400)$ observed in tau decay 
can further constrain $\theta_K$ and yield the size of $SU(3)_f$ 
symmetry-breaking.

\begin{figure}[htb]
\vskip1.7in
\caption{$\tau^-\rarrow 
\bar{K^0}\pi^-\pizero\nu_\tau$ branching fractions.}
\label{fig:fk0pipi0nu}
\includegraphics{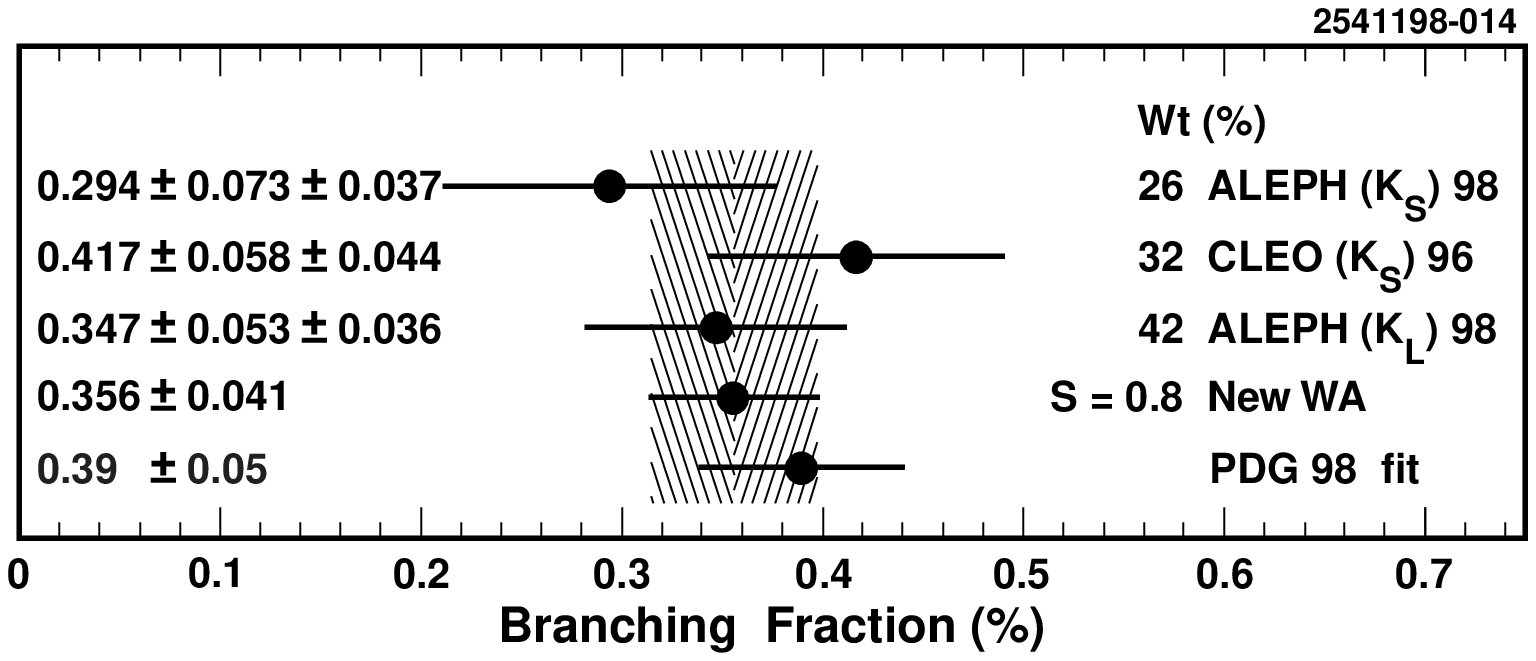}
\end{figure}

\begin{figure}[htb]
\vskip1.2in
\caption{$\tau^-\rarrow K^-\pi^+\pi^-\nu_\tau$ branching fractions.}
\label{fig:fk2pinu}
\includegraphics{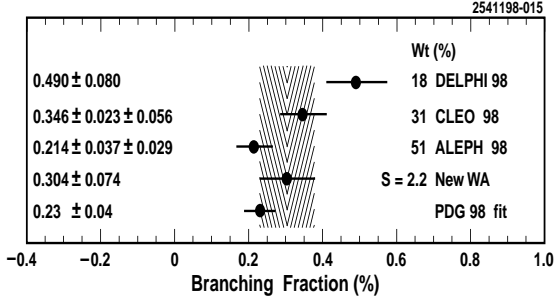}
\end{figure}

   ALEPH \cite{Chen,ALEPHKaon1pr,ALEPHKaonSpc} presented new information
on all four possible $K\pi\pi$ final states, while CLEO
\cite{Kravchenko} and DELPHI \cite{Andreazza}
did so for the all-charged mode. The updated ALEPH value 
$\BR{K^-2\pizero}$=(0.056\qm0.025)\% is now corrected for
$K^-3\pizero$ feed-down. Fig.~\ref{fig:fk0pipi0nu}
compare the CLEO and ALEPH results for the $K^0$ modes, which are 
consistent, and Fig.~\ref{fig:fk2pinu} does the same for the
$K^-\pi^+\pi^-$ mode, for which the three results differ
substantially from one another. The theoretical prediction
from Li \cite{Li} of 0.18\% is closer to the ALEPH and
CLEO result than Finkemeier and Mirkes \cite{Fink} prediction of 0.77\%,
which is much higher even than the DELPHI measurement.

  CLEO \cite{Kravchenko} and ALEPH \cite{Chen,ALEPHKaonSpc} have both
made quantitative analyses of the resonance structure of $(K\pi\pi)^-$ 
decays. CLEO has $\sim$8K events in $K^-\pi^+\pi^-$ with 38\% background
and is systematics-limited, while ALEPH has $\sim$300 events split
between the $K^-\pi^+\pi^-$ and $\bar{K^0}\pi^-\pizero$ modes, each with
$\sim$50\% background. CLEO fits the sub-mass spectra for $K_1(1270)$, 
$K_1(1400)$ amplitudes, strong phases in their $K^-\rho$ and $K^{*0}\pi^-$ 
decays, and the mixing angle $\theta_K$. Accounting for the interference
is found to be important, as the two phases are about 5 and 1.7 standard
deviations from zero: the predicted mass spectra match the data
well only if these phases are included. The $K_1(1270)$ is found to
dominate, and CLEO quotes the ratio of the two $K_1$ states as 
$K_1(1270)$/$K_1(1400)$=4.1\qm0.5$^{+4.8}_{-2.1}$.
The final state $K\rho$ is found to be (45\qm12)\% of $K^-\pi^+\pi^-$.
Finally, the mixing angle is determined within a two-fold
ambiguity to be (67\deg~or~46\deg)\qm10\deg, close to the 
values obtained from $K_1$ widths and branching fractions \cite{Suzuki93}
of (57\deg~or~33\deg).

  In contrast, ALEPH assumes the $K\pi\pi$ state appears as an 
{\it incoherent} superposition of $K\rho$ and $K^*\pi$, finding
$K\rho$ to be (35\qm11)\% of $K^-\pi^+\pi^-$ and~(66\qm12)\% of
$\bar{K^0}\pi^-\pizero$ (consistent with isospin symmetry),
the fraction of $K_1(1270)$ in the final states to be $(41\pm19\pm15)$\%,
and the branching fraction $\BR{K_1(1270)}=(0.45\pm0.11)$\%.
So, while ALEPH and CLEO use different assumptions and analyze
their data differently, qualitatively, they agree that $K\rho$ is
a significant presence in  $K\pi\pi$ decays, and therefore that
$K_1(1270)$ plays a major role as a source of these final states.

 Another final state may proceed through the $K_1$ resonance.
CLEO \cite{CLEOKsteta} has found
$\BR{K^{*-}\eta}$=(2.9\qm0.8\qm0.4)$\times 10^{-4}$, finding
that the $K^*$ is dominant in $(K\pi)^-\eta$ decays. Li \cite{Li}
predicts that this final state proceeds through the $K_1$ 
axial-vector current, and, using an effective chiral theory in
the limit of chiral symmetry, predicts a branching fraction of
(1.01$\times 10^{-4}$). In contrast, Pich \cite{Pich} predicts
a rate for $(K\pi)^-\eta$ of $\sim10^{-5}$ using chiral perturbation theory.

\subsection{$\tau\rarrow (K\pi\pi\pi)^-\nu_\tau$}

\begin{figure}[htb]
\vskip1.3in
\caption{$\tau^-\rarrow K^-\pi^+\pi^-\pi^0\nu_\tau$ branching fractions.}
\label{fig:fk2pipi0nu}
\includegraphics{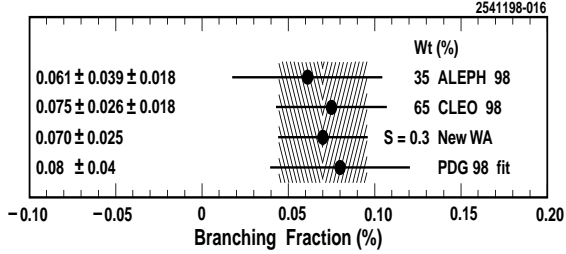}
\end{figure}

 ALEPH \cite{Chen,ALEPHKaonSpc} and CLEO \cite{Kravchenko} both look for
$(K3\pi)^-$ final states, do not obtain definitive signals for any 
of them, but determine sub-0.1\% branching fractions. Fig.~\ref{fig:fk2pipi0nu}
shows the measurements for $K^-\pi^+\pi^-\pi^0$, which are consistent
with each other. ALEPH also measures 
$\BR{\bar{K^0}\pi^-2\pizero(excl.~K_S)}=(0.058\pm0.036)$\%,
$\BR{K^-3\pizero(excl.~\eta)}=(0.037\pm0.024)$\%,
and $\BR{K^03h^\pm(excl.~K_S)}=(0.023\pm0.020)$\%.

\begin{figure}[tbh]
\vskip1.2in
\caption{Branching fractions for $\tau^-\rarrow K^0 K^-\nu_\tau$.}
\label{fig:fkk0nu}
\includegraphics{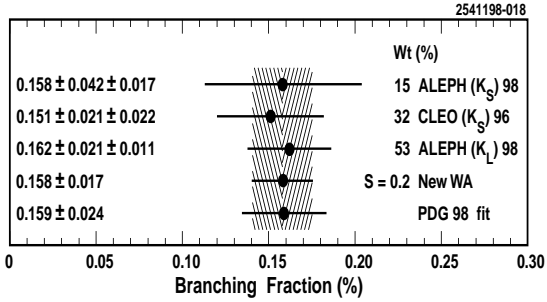}
\end{figure}

\subsection{Total Strange Spectral Function}

  ALEPH \cite{Hoecker} has combined the spectral information from
$K\pi$ and $K\pi\pi$ decays with small additions from Monte Carlo
for $K\ge3\pi$ modes to determine the total strange spectral function.
Using this and the strange-quark mass dependence of the
QCD expansion for the hadronic decay width of the tau, ALEPH
obtains $m_S$(1~GeV$^2$)$=(217^{+45}_{-57})$~MeV.

\begin{figure}[htb]
\vskip1.8in
\caption{$\tau^-\rarrow K^-K^+\pi^-\nu_\tau$ branching fractions.}
\label{fig:fkkpinu}
\includegraphics{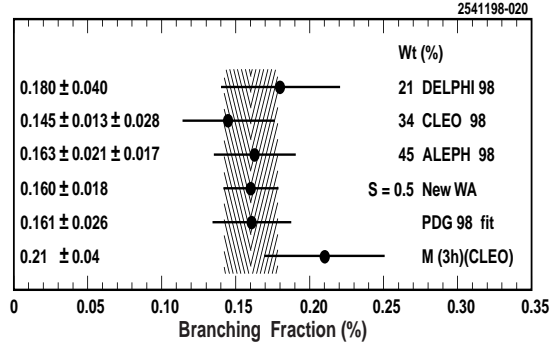}
\end{figure}

\begin{figure}[htb]
\vskip0.85in
\caption{$\tau^-\rarrow K_S^0\bar{K_S^0}\pi^-\nu_\tau$ branching fractions.}
\label{fig:fkskspinu}
\includegraphics{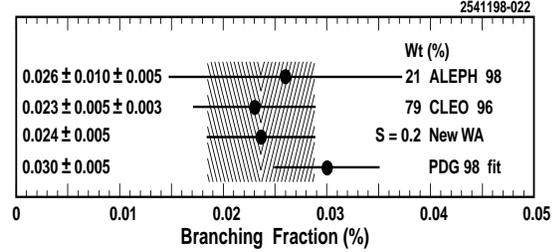}
\end{figure}


\begin{figure}[htb]
\vskip1.65in
\caption{$\tau^-\rarrow K^0 K^-\pizero\nu_\tau$ branching fractions.}
\label{fig:fk0kpi0nu}
\includegraphics{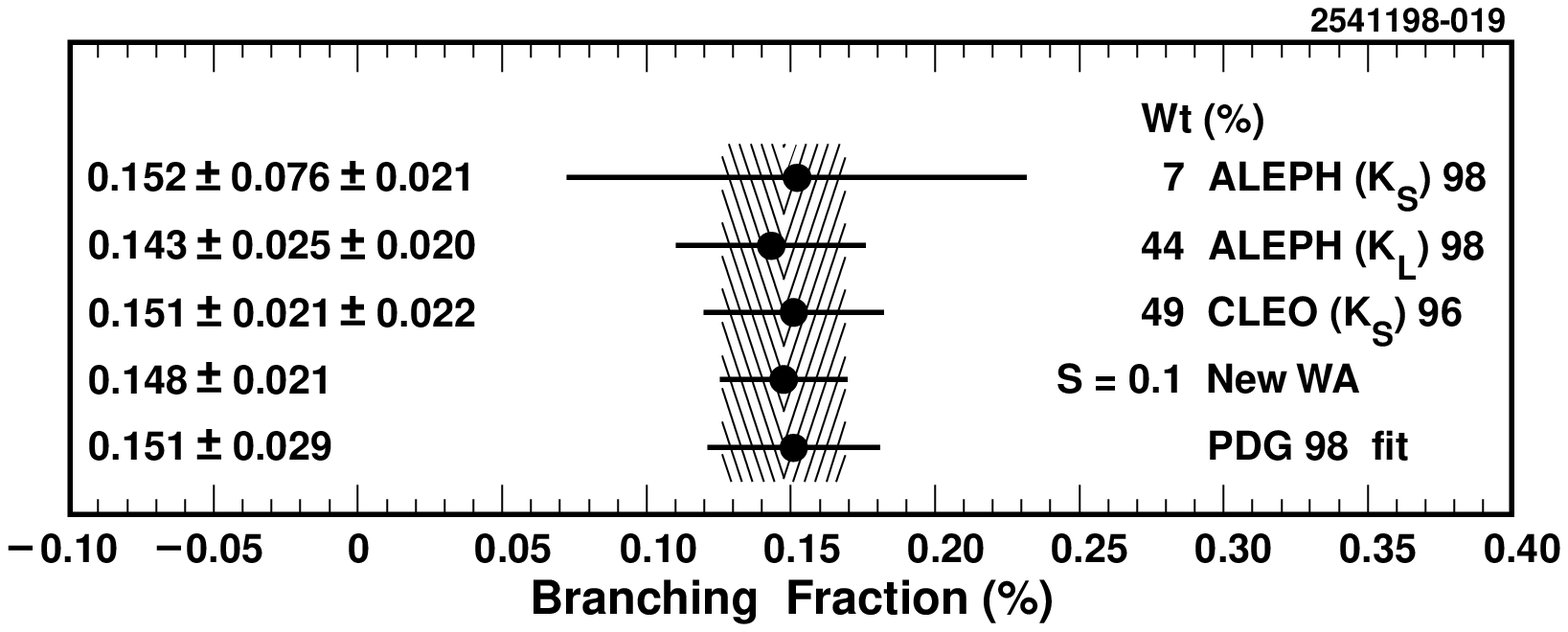}
\end{figure}

\begin{figure}[htb]
\vskip1.1in
\caption{$\tau^-\rarrow K^-K^+\pi^-\pi^0\nu_\tau$ branching fractions.}
\label{fig:fkkpipi0nu}
\includegraphics{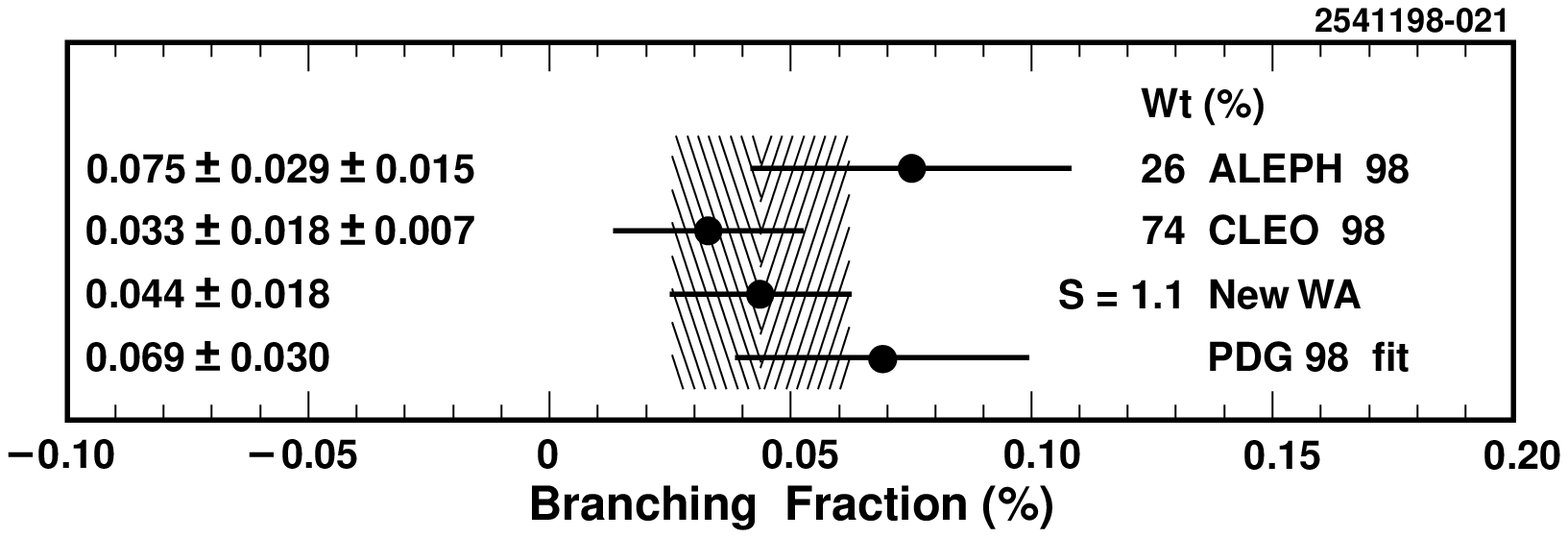}
\end{figure}

\subsection{$\tau\rarrow (KK)^-\nu_\tau$,
$(KK\pi)^-\nu_\tau$,  $(KK\pi\pi)^-\nu_\tau$
}

    The production of $(KK)^-$ or $(KK\pi)^-$ states in tau decay is suppressed
by phase space, but could occur via direct fragmentation or through the 
high-mass tails of the $\rho^-$, $a_1^-$, or $\rho^{\prime -}$:
$\rho^-\to (KK)^-$, $\rho^{\prime -}\to (K^*K)^-$, 
$a_1^-\to(\rho\pi)^-\to (KK\pi)^-$, or $a_1^-\to(K^*K)^-$.  
ALEPH and CLEO have new results in many of these modes, including 
$K^-K^+\pi^-\pizero$, as shown in Figs.~\ref{fig:fkk0nu}-\ref{fig:fkkpipi0nu}. 
There are no major inconsistencies in the measured branching fractions, which 
are typically $\sim$0.1\%. The observation by CLEO of $K^*K$
content in $3h^\pm$ events is quantitatively consistent with the
measured branching ratio: that analysis \cite{Schmidtler,CLEO3pi} determines 
${\cal B}(\tau\to K^*K\nu_\tau)/{\cal B}(\tau\to 
(3h)^-\nu_\tau)$=(3.3\qm0.5\qm0.1)\%, which results in the prediction for
the $K^-K^+\pi^-$ branching fraction shown in Fig.~\ref{fig:fkkpinu}.

The ALEPH spectrum for $K_SK^-$ does not match the model expectation for
$\rho\to K\bar{K}$, in agreement with the conclusion of CLEO \cite{CLEOKsh}. 
The ALEPH analysis also favors $K^*K$ parentage over $\rho\pi$ in $KK\pi$ 
states, and quotes an axial-vector current fraction of
(94$^{+6}_{-8}$)\% based on CVC and the absence of $KK\pi$ in
$e^+e^-$ data.

 Some comments on the $K^0\bar{K^0}\pi^-$ final state \cite{PDG98} are 
in order. ALEPH sees 68 $K_SK_L\pi^-$ events and 6 in $K_SK_S\pi^-$,
compared to CLEO's 52 events \cite{CLEOKsh} in the latter mode only. 
If the $K^0$ and $\bar{K^0}$ decay independently, the  $K_SK_L\pi^-$ mode
should have double the rate of $K_SK_S\pi^-$ or $K_LK_L\pi^-$. However,
Bose-Einstein correlations between the neutral kaons will violate this 
assumption. The ALEPH measurement \cite{Chen}
of $\BR{K_SK_L\pi^-}=(0.101\pm0.023\pm0.013)\%$ is two
standard deviations higher than double the NWA $\BR{K_SK_S\pi^-}$,
suggestive that such an effect may be in force, but errors are
currently too high for any definitive conclusion.

\section{Conclusions}

  Some perspective on recent experimental accomplishments can be gained by
answering some of the questions I posed in the TAU94 
proceedings \cite{Heltsley94}:
{\sl Will the global branching fractions of ALEPH continue to dominate 
most modes, or can we expect comparably precise results from other 
experiments?}
ALEPH still sets the experimental standard for comprehensive treatment 
of hadronic tau decays, due to a combination of the clean tau-pair 
topology at LEP energies, a superbly designed and maintained detector, 
hard work, and aggressive analysis.
This is particularly true of decays involving a \KL, or a charged kaon
{\sl and} \piz's, or modes with multiple \piz's. CLEO competes best
on modes with \piz's and/or with branching fractions below 1\% where
its large dataset can provide enough events. In a very welcome
development, OPAL and DELPHI have
recently achieved better understandings of their detector systematics
and are pressing or even surpassing ALEPH and CLEO on selected modes.
It is a trend we hope can continue. However, we all await ALEPH's
final full-data-sample global analysis with trepidation!

{\sl Can we resolve branching fraction discrepancies in the $h\pizero$
and $3h$ modes simply by improving measurements, making older,
disagreeable values irrelevant? } Apparently so.

{\sl Can we develop procedures which allow a consistent accounting of 
modes with \KL\ decays?} PDG96 \cite{PDG96} adopted basis modes
which excluded unseen \KL, and measurements of modes with
\KL, dominated by ALEPH \cite{Chen},
have made the feed-across corrections feasible.

{\sl What are the masses and widths of the $\rho$ and $\rho'$ in tau
decay?} ALEPH results have been published \cite{ALEPHrho},
but the preliminary results from CLEO \cite{Urheim96} have
not.

{\sl Is all $(3\pi)^-$ attributable to $a_1^-$ decays? }
A poorly-phrased question; better would have been ``are there
intermediate states in  $(3\pi)^-$ that do not go through  $a_1\to\rho\pi$?''.
The CLEO analysis \cite{Schmidtler,CLEO3pi} shows that less
than 70\% of $\pi^-\pi^0\pi^0$ final states come from $\rho\pi$
intermediate states, and in particular that isoscalars 
($\sigma\pi$, $f_0(1186)\pi)$ amount to $\sim$20\% of the branching
fraction. The CLEO and DELPHI \cite{DELPHI3pi} analyses also
are suggestive of the presence of an $a_1^\prime$ at high $3\pi$ mass.

{\sl Is there evidence for scalar or non-resonant production of 
$(K\pi)^-$ final states? }
No, but statistics are scarce for this study. I should have asked
``is there evidence for $K^*(1410)$ in $K\pi$ decays'', to which
the answer is, yes, from ALEPH \cite{Chen,ALEPHKaonSpc}.

{\sl Are $(K\pi\pi)^-$ decays consistent with a mixture of $K_1(1270)$ and 
$K_1(1400)$ production?} Yes, and both CLEO \cite{Kravchenko} and 
ALEPH \cite{Chen} agree that $K_1(1270)$ amounts to a substantial fraction 
of the $(K\pi\pi)^-$ final state.
{\sl  What is the mixing angle and strength of $SU(3)_f$ symmetry breaking 
in such decays?} CLEO has a preliminary analysis \cite{Kravchenko}, which 
accounts for interference between the different intermediate states, 
in which the mixing angle is measured up to a two-fold ambiguity.
{\sl  Do $(K\pi\pi)^-$ final states have $K^*(1410)$ or non-resonant
content?} ALEPH \cite{Chen,ALEPHKaonSpc} has measured 
$\BR{K^*(1410)}$=(0.15$^{+0.14}_{-0.10}$)\%,
and so the branching fraction into $(K\pi\pi)^-$ can be computed. 
Statistics are too low to answer for non-resonant
production.

{\sl Can all $(4\pi)^-$ final states be described as 
$\omega\pi^-$ and $(a_1\pi)^-$? If so, what resonances produce those
states? } A complete spectral analysis of $\tau^-\to(4\pi)^-\nu_\tau$ 
has yet to be done.

{\sl What is the resonant substructure in $KK\pi$, $5\pi$,
$6\pi$ final states?} The $KK\pi$ final state seems to arise
through the axial-vector current $a_1^-\to K^*K$ \cite{Chen,ALEPHKaonSpc}.
Five and six pion final states are quite complex, but receive
important contributions from $f_1\pi$ and 
$\pi^-2\pizero\omega$ \cite{CLEOf1pi,Gan,CLEO3pi3pi0}.

{\sl What is the rate for 7$\pi$ production in tau decay?}
For seven charged pions with or without \piz's, less than 2.4 parts 
per million \cite{Gan,CLEO7pi}. No direct measurement of $3\pi^\pm4\pizero$
has been performed and only upper limits exist for $5\pi^\pm2\pizero$.

{\sl Is there any evidence for second-class currents in tau decay at 
expected or unexpected levels?} Sadly, nothing unexpected, and
the expected levels have yet to be approached experimentally.

{\sl Given the current status of measurements, where are the most promising
places to look for deviations from the Standard Model in hadronic tau
decays?}
Recent effort has been focussed on rare, suppressed, or
forbidden decays \cite{Gan}, and on CP-violation \cite{Kass},
but any Standard Model prediction is fair game.

  Looking ahead, as the LEP and CLEO~II programs draw to a close, 
finalized, full-statistics data-sets should become available next year 
for analysis. Improved low-energy data for CVC tests should continue
to emerge from the experiments at VEPP-2M. Considerably more hadronic 
tau physics awaits further mining of all these samples, both with 
high-statistics systematics-limited studies which need more time and
data to mature, and for low-statistics searches which can benefit the 
most from additional luminosity. Experimenters should not be dissuaded 
from this pursuit by the somewhat disquieting over-consistency of many 
results; surprise often lurks just beyond the horizon. We should all
look forward to a peek at something new at TAU2000.

\section{Acknowledgements}

     I would like to thank other speakers for making their
results available to me for the writing of this summary talk. In particular, 
I benefitted greatly from communications with S.~Chen, M.~Davier, S.~Eidelman,
J.M.~Lopez-Garcia, and R.~Sobie. My fellow CLEOns continue
to indulge my endless questions, especially J.~Smith, 
I.~Kravchenko, D.~Besson, and A.~Weinstein. The Organizers and 
Secretariat are to be commended for hosting a productive
and invigorating Workshop. The preparation of this manuscript 
was supported by the National Science Foundation.

\def\WKSHP{presentation at this Workshop}

\end{document}